\documentclass{conm-p-l}
\usepackage{amssymb}
\begin{document}

\newtheorem{theorem}{Theorem}[section]
\newtheorem{lemma}[theorem]{Lemma}

\theoremstyle{definition}
\newtheorem{definition}[theorem]{Definition}
\newtheorem{example}[theorem]{Example}
\newtheorem{xca}[theorem]{Exercise}
\newtheorem*{theorem*}{Theorem}

\theoremstyle{remark}
\newtheorem{remark}[theorem]{Remark}

\numberwithin{equation}{section}

\newcommand{\re}{\operatorname{Re}}
\newcommand{\dv}{\operatorname{div}}
\newcommand{\rsp}{\operatorname{rsp}}
\newcommand{\curl}{\operatorname{curl}}
\newcommand{\ess}{\operatorname{ess}}
\newcommand{\supp}{\operatorname{supp}}
\newcommand{\comp}{\operatorname{comp}}

\title[Essential Spectrum of the Linearized 2D
Euler Operator]{The Essential Spectrum of the Linearized 2D
Euler Operator is a Vertical Band}

\author{Roman Shvidkoy \and Yuri Latushkin}
\address{Department of Mathematics, University of Missouri, Columbia, MO 65211}
\email{shvidkoy@math.missouri.edu, yuri@math.missouri.edu}
\thanks{The second author was partially
supported by the Twinning Program of the
 National Academy of Sciences
and National Science Foundation, and by the
 Research Council and Research Board
of the University of Missouri}%
\date{\today}

\subjclass{76,35}

\keywords{Inviscid fluids, Euler equations,  essential
spectrum}
%\maketitle

\begin{abstract} We prove that the essential spectrum of the operator obtained by linearization about a steady state of the
Euler equations governing the motion of inviscid ideal fluid in
dimension two is a vertical strip whose width is determined by the
maximal Lyapunov exponent of the flow induced by the steady
state.\end{abstract} \maketitle
\section{Introduction}

In this note we continue the work in \cite{SL}, and give a full
description of the essential spectrum for the linearized Euler
operator $L$ in dimension two. We prove that the essential
spectrum of the operator is one solid vertical strip symmetric
with respect to the imaginary axis. The width of the strip is
determined by the maximal Lyapunov exponent $\Lambda$ for the flow
induced by the steady state.

For classical results concerning linearized Euler equations see,
e.g., \cite{C,DR,L,Y}. Recent advances concerning the essential
spectrum of the linearized Euler operator can be found in
\cite{FSV,FSV2,FV,FV2,V,VF}. In particular, it was proved in \cite{V,VF}
that the essential
growth bound for the group generated by $L$ is equal to $\Lambda$. Using this
result, it was proved in \cite{LV} that the essential spectral bound for $L$ is equal to
$\Lambda$.  

We study the linearized
Euler operator $L$ in vorticity form,
$$Lw=-\langle u,\nabla \rangle w-\langle \curl ^{-1} w,
\nabla \rangle \curl u,$$
on the Sobolev space $H^0_1=H^0_1(\mathbb{T}^2)$ of scalar functions $w$ having zero means $\int wdx=0$ on the $2$-torus
$\mathbb{T}^2=\mathbb{R}^2 /2\pi \mathbb{Z}^2$. Here $u=(u_1,u_2)^\top$ is a steady state (velocity) solution of the Euler
equations $\langle u,\nabla \rangle u+\nabla p=0$, $\dv u=0$, $p$ is the pressure, $\langle \cdot ,\cdot \rangle$ denotes the
scalar product, $\top$ -- transposition,  $\curl u=-\partial _2u_1+\partial _1u_2$ is the scalar curl of the two-dimensional vector field, and
$v=\curl ^{-1}w$ denotes the unique solution of $\curl v=w$, $\dv v=0$.

Observe, that in the representation $L=-A+T$, $Aw=\langle u,\nabla
\rangle w$, $Tw=-\langle \curl ^{-1}w,\nabla \rangle \curl u$, the
operator $T$ is compact. The operator $A$ generates {\em the
evolution semigroup} $e^{tA}w(x)=w\circ \varphi_t(x)$, cf.
\cite{CL}. Here $\varphi_t:x_0\mapsto x(t;x_0)$ is the flow on
$\mathbb{T}^2$ generated by the steady state $u$, that is, by the
solutions of the equation $\partial_tx(t)=u(x(t))$.

Our approach is to describe
 $\sigma_{ap}(L)$, the approximative point
spectrum of $L$, by looking for  a {\em weakly null
approximative eigenfunction} for $A$, that is, for an
$\alpha=\lambda+i\xi\in{\mathbb{C}}$ and for a sequence
$\{f_n\}\subset H^0_1$ such that $\| f_n\|_{H^0_1}=1$,
$\lim_{n\to \infty}\| (A-\alpha)f_n\|_{H^0_1}=0$, and $\{ f_n\}$
converges to $0$ weakly in $H_1^0$ 
as $n\to \infty$. As soon as this sequence is
 found, $(L+\alpha )f_n=-(A-\alpha )f_n+Tf_n\to 0$
 in $H^0_1$ as $n\to \infty$, and hence, $-\alpha \in \sigma _{ap}(L)$.

We modify a construction presented in \cite{SL}, and find weakly
null approximative eigenfunctions for $A$ supported in thin and
long flow boxes around stable and unstable orbits for the
hyperbolic stagnation point $y$ where the maximal Lyapunov
exponent $\Lambda$ is attained. To describe our construction, let
$u^\bot=(-u_2,u_1)^\top$ and denote by $\{ \psi_t\}$ the flow on
$\mathbb{T}^2$ induced by $u^\bot$. Fix $x_0=x_1$ on the stable
orbit (respectively, $x_0=x_2$ on the unstable orbit) of $y$.
Define a bijection $H(t,\tau )=\varphi_t \circ \psi_\tau(x_0)$ on
a strip $\mathcal{S}=\{(t,\tau)\in\mathbb{R}^2 : |t|\leq N$ and
$|\tau |\leq s\}$, where $s$ is sufficiently small and $N$ is
arbitrarily large. Note that under the transformation $H$ the
operator $A$ becomes simply $\partial_t$. For
$\alpha=\lambda+i\xi\in \mathbb{C}$ let
\begin{equation}\label{DefF} F(t,\tau)=e^{\alpha t}\gamma (t)\beta (\tau ),\quad (t,\tau)\in \mathcal{S},\end{equation}
where smooth cut-off functions $\beta$ and $\gamma$ are appropriately
chosen such that
$\supp \beta \subset (-s,s)$ and $\supp \gamma \subset (-N,N)$. Put $f=F\circ H^{-1}$. Then, by a direct computation,
we have:
\begin{equation}\label{DefFTil} Af-\alpha f\mid_{H(t,\tau )}=
\tilde{F}(t,\tau ),\quad\text{where } \tilde{F}(t,\tau
):=e^{\alpha t}\gamma'(t)\beta (\tau ).\end{equation}
Note that if $s\to 0$ then $f\to 0$ weakly.
For $\lambda \in (-\Lambda ,\Lambda )$ we will choose below
 a sequence of functions
$\gamma=\gamma_{K,M}$ such that if $K,M\to \infty$ then $\| \tilde{F}\circ H^{-1}\|_{H_1^0}/\| F\circ H^{-1}\|_{H_1^0}\to 0$. This
shows that $\alpha \in \sigma_{ap}(A)$ and $-\alpha \in \sigma_{ap}(L)$.

We use the following notations: For an operator $B$ on $H^0_1$ we
denote by $\sigma (B)$ and $\sigma _{\ess }(B)$ the spectrum and
essential (Weyl) spectrum, by $\rsp (B)$ and $\rsp \ess (B)$ the
spectral radius and essential spectral radius, by $\omega
(B)=t^{-1}\log \rsp (e^{tB})$ and $\omega _{\ess }(B)=t^{-1}\log
\rsp \ess (e^{tB})$ the growth and essential growth bound of the
semigroup $\{ e^{tB}\}$, by $s(B)=\sup \{\re \alpha :\alpha \in
\sigma (B)\}$ the spectral bound. For $\alpha =\lambda +i\xi \in
\mathbb{C}$ we denote $\| L+\alpha \|_\bullet=\inf \{ \|
(L+\alpha)g\|_{H_1}:\| g\|_{H_1}=1\}$. We write $a\lesssim b$ if
$a\leq cb$ for a constant $c$ independent of $a$ and $b$.

\bigskip

\noindent{\bf Acknowledgments.} The authors are grateful to Susan 
Friedlander and Misha Vishik for many discussions.

\section{Results}

Let $\Sigma$ denote the set of all Lyapunov exponents for the differential $\{ D\varphi _t\}$,
given by the Multiplicative Ergodic Theorem~\cite{O}, and $\Lambda =\max\{ \lambda :\lambda \in \Sigma \}$ denote the maximal
exponent:
$$\Lambda =\lim_{t\to \infty} \max_{x\in \mathbb{T}^2}t^{-1}\log \| D\varphi_t(x)\|.$$

\begin{remark}\label{rem1} Since $\dv u=0$, if $y$ is a hyperbolic
stagnation point for $u$, then  $\sigma
(Du(y))=\{ -\lambda ,\lambda \}$, where $\lambda >0$ is the Lyapunov  exponent
for $\{ D\varphi_t\}$ at $y$. Also, if $\lambda
\in \Sigma \backslash \{0\}$ then there exists a hyperbolic stagnation point $y$ such that $\lambda$ is the Lyapunov exponent
for $\{D\varphi _t\}$ at $y$, see \cite[Rem. 6]{SL}.\end{remark}

Let $p(x)=\inf \{ t>0:\varphi_t(x)=x\}$ denote the prime period of
$x\in \mathbb{T}^2$. We set $p(x)=\infty$ if the point $x$ is
nonperiodic. We say that the flow $\{ \varphi_t\}_{t\in
\mathbb{R}}$ has {\em arbitrarily long trajectories} if for each
$N\in \mathbb{N}$ there is an $x\in \mathbb{T}^2$ such that
$p(x)\geq N$.

The following facts have been proved in \cite{SL}:
\begin{enumerate}
\item[(i)] if $\lambda \in \Sigma \backslash \{0\}$ then $\lambda
+i\mathbb{R}\subset \sigma_{\ess}(L)$;
\item[(ii)] $\omega_{\ess}(L)=\omega _{\ess}(-L)=\Lambda$;
\item[(iii)] $s(L)=\omega (L)$;
\item[(iv)] if
$\{ \varphi _t\}$ has arbitrarily long trajectories then
$i\mathbb{R}\subset \sigma_{\ess}(L)$; if, in addition, $\Lambda =
0$ then $i\mathbb{R} =\sigma_{\ess}(L)$.
\end{enumerate}

In this paper we complete the
description of $\sigma_{\ess}(L)$ and $\sigma _{\ess}(e^{tL})$, $t\in \mathbb{R}$, as follows.

\begin{theorem*} If $\Lambda >0$ then $\sigma_{\ess}(L)=\{\alpha \in \mathbb{C}:|\re \alpha|\leq \Lambda \}$ and
$\sigma_{\ess}(e^{tL})=\{ z\in \mathbb{C}:
e^{-|t|\Lambda}\leq |z|\leq e^{|t|\Lambda }\}$,
$t\in \mathbb{R}$.\end{theorem*}

\begin{remark} All facts listed above and the Theorem hold true if $\sigma_{\ess}(L)$ is replaced by $\sigma(A)$. For higher regularity
Sobolev spaces $H^0_m$, $m\in \mathbb{Z}$,
one can prove similar assertions replacing $\lambda $ by $m\lambda$ and $\Lambda $
by $|m|\Lambda$.\end{remark}

The main step in the proof of the Theorem is
the following proposition from \cite{SL} (see (10) in the proof of Theorems~1 and 2
there); for completness we sketch its proof in Appendix.

\begin{lemma}\label{LemSL} If $\alpha =\lambda +i\xi$, $\lambda$,
$\xi \in \mathbb{R}$, $x_0\in \mathbb{T}^2$,
$N<p(x_0)/2$, and $\gamma\in H_1([-N,N])$
is a real valued cut-off function with
$\supp \gamma \subset (-N,N)$,  then
\begin{equation}\label{Old10} \| L+\alpha \|^2_\bullet
\leq \frac{\int\limits_{\mathbb{R}}|u\circ
\varphi_t(x_0)|^2/|u(x_0)|^2e^{2\lambda t}|\gamma'(t)|^2dt}{\int
\limits_{\mathbb{R}}|u\circ
\varphi_t(x_0)|^2/|u(x_0)|^2e^{2\lambda t}|\gamma (t)|^2dt}.\end{equation}\end{lemma}

\begin{proof}[Proof of Theorem] We prove that $(-\Lambda ,\Lambda)\subset \sigma_{ap}(L)$. The rest follows from the Spectral Inclusion Theorem
$\exp t\sigma (L)\subset \sigma (e^{tL})$ and fact (ii) above.

Using Remark~\ref{rem1}, pick the hyperbolic stagnation point $y$ such that $\sigma (Du(y))=\{-\Lambda ,\Lambda \}$.
Fix
$\lambda \in (0,\Lambda)$ and consider any point $x_1$, $u(x_1)\neq 0$, that belongs to the orbit attracted to $y$ such that
$y=\lim_{t\to \infty}\varphi_t(x_1)$. Respectively, fix $\lambda \in (-\Lambda ,0)$ and consider any point $x_2$, $u(x_2)\neq
0$, that belongs to the orbit repelled from $y$ such that $y=\lim_{t\to -\infty}\varphi _t(x_2)$.

\begin{lemma}\label{EstExp} There exists $\epsilon =\epsilon (\lambda )>0$ such that
\begin{gather}\label{EstF} \sup\{e^{\epsilon t}e^{\lambda t}
|u\circ\varphi _t(x_1)|:t\geq 0\}<\infty;\\
\label{EstB}\sup\{ e^{-\epsilon t}e^{-\lambda t}|u\circ\varphi_t(x_2)|:t\leq
0\}<\infty.\end{gather}\end{lemma}

The proof of the lemma is given in Appendix.

Fix $K>0$, $M>0$. We will construct a sequence of cut-off functions $\gamma=\gamma_{K,M}\in
H_1([-N,N])$, $N>K+2M$,
such that for all $\xi \in \mathbb{R}$ the right hand side of \eqref{Old10} tends to zero as $K,M\to \infty$. By
Lemma~\ref{LemSL} this implies $\alpha =\lambda +i\xi \in \sigma_{ap}(-L)$ which proves the theorem.

For $\lambda \in (0,\Lambda )$ the function $\gamma$ is defined as follows: If $t\leq 0$ then $\gamma (t)=0$; if $0<t\leq K$,
then $\gamma (t)=t/K$; if $K<t\leq K+M$ then $\gamma (t)=\exp (-(t-K)/M)$; if $K+M<t\leq K+2M$ then $\gamma
(t)=-e^{-1}(t-K)/M+2e^{-1}$; if $t>K+2M$ then $\gamma (t)=0$. Denote
$x_0=\varphi _{-K}(x_1)$, and observe that the right-hand
side of \eqref{Old10} can be estimated from above by the following expression:
$$\dfrac{\left( \frac{1}{K^{2}}\int\limits^K_0+\frac{1}{e^{2}M^{2}}
\int\limits^{K+2M}_{K+M}\right) |u\circ
\varphi_t(x_0)|^2e^{2\lambda
t}dt+\frac{1}{M^{2}}\int\limits^{K+M}_Ke^{-\frac{2(t
-K)}{M}}|u\circ \varphi_t(x_0)|^2e^{2\lambda
t}dt}{\int\limits^{K+M}_Ke^{-\frac{2(t-K)}{M}}|u\circ
\varphi_t(x_0)|^2e^{2\lambda t}dt}.$$ By \eqref{EstF}, $e^{\lambda
t}|u\circ\varphi_t(x_0)|\lesssim e^{-\epsilon t}$ uniformly for
$t\geq 0$. Therefore, we can pass to the limit as $M\to \infty$:
\begin{equation*}\begin{split} \| L+\alpha \|^2_\bullet
&\leq \frac{K^{-2}\int\limits^K_0|u\circ \varphi_{t-K}(x_1)|^2e^{2\lambda
t}dt}{\int\limits^\infty_K|u\circ \varphi_{t-K}(x_1)|^2e^{2\lambda t}dt}\\
&=\frac{K^{-2}e^{2\lambda K}\int\limits^0_{-K}|u\circ \varphi_t(x_1)|^2e^{2\lambda t}dt}{e^{2\lambda K}
\int\limits^\infty_0|u\circ
\varphi_t(x_1)|^2e^{2\lambda t}dt}\\
&\leq \frac{\| u\|_\infty}{K^2}\frac{\int\limits^0_{-\infty}e^{2\lambda
t}dt}{\int\limits^\infty_0|u\circ \varphi_t(x_1)|^2e^{2\lambda
t}dt}.\end{split}\end{equation*}
Note that the improper integrals converge due to \eqref{EstF} and $\lambda >0$. Letting $K\to \infty$, we have $-\alpha \in
\sigma_{ap}(L)$.

For $\lambda \in (-\Lambda ,0)$ the function $\gamma$ is defined
as follows. If $t\geq 0$ then $\gamma (t)=0$, if $-K\leq t<0$ then
$\gamma (t)=-t/K$, if $-(K+M)\leq t<-K$ then $\gamma (t)=\exp
((t+K)/M)$; if $-(K+2M)\leq t<-(K+M)$ then $\gamma
(t)=e^{-1}(t+K)/M+2e^{-1}$; and if $t<-(K+2M)$ then $\gamma
(t)=0$.
 Denote $x_0=\varphi_K(x_2)$. Then the right-hand side of
\eqref{Old10} can be estimated from above by the following expression:
\begin{equation*}\begin{split} &\quad
\dfrac{\int\limits^0_{-\infty}|u\circ\varphi_t(x_0)|^2e^{2\lambda
t}|\gamma'(t)|^2dt}{\int\limits^0_{-\infty}|u\circ\varphi_t(x_0)|^2
e^{2\lambda t}|\gamma (t)|^2dt}\le\\
& \dfrac{\Big( \frac{1}{e^{2}M^{2}}
\int\limits^{-(K+M)}_{-(K+2M)}+\frac{1}{K^{2}}
\int\limits^0_{-K}\Big) |u\circ \varphi_t(x_0)|^2e^{2\lambda
t}dt+\frac{1}{M^{2}}
\int\limits^{-K}_{-(K+M)}e^{\frac{2(t+K)}{M}}|u\circ
\varphi_t(x_0)|^2e^{2\lambda
t}dt}{\int\limits^{-K}_{-(K+M)}e^{\frac{2(t+K)}{M}}|u\circ \varphi_t(x_0)|^2e^{2\lambda t}dt} .\end{split}\end{equation*}
By \eqref{EstB}, $e^{\lambda t}|u\circ \varphi_t(x_0)|\lesssim e^{\epsilon t}$ uniformly for $t\leq 0$. Similarly to the case
$\lambda \in (0,\Lambda )$, we conclude that $-\alpha \in \sigma_{ap}(L)$.
\end{proof}

\section*{Appendix}

\begin{proof}[Proof of Lemma~\ref{EstExp}] Choose $v_{1,2}\in \mathbb{R}^2$ such that $|v_1|=|v_2|=1$, and $Du(y)v_1=-\Lambda
v_1$, $Du(y)v_2=\Lambda v_2$. Then $D\varphi _t(y)=e^{tDu(y)}$ and
$$\lim_{t\to +\infty}\frac{u\circ\varphi _t(x_1)}{|u\circ\varphi_t(x_1)|}=v_1,\quad \lim_{t\to
-\infty}\frac{u\circ\varphi_t(x_2)}{|u\circ\varphi_t(x_2)|}=v_2$$
imply:
\begin{equation}\begin{split}\label{LambdaEst} \lim_{t\to +\infty}\log \left|
D\varphi
(\varphi_t(x_1))\frac{u\circ\varphi_t(x_1)}{|u\circ\varphi _t(x_1)|}\right|
 & =-\Lambda ,\\
\lim_{t\to -\infty}\log \left| D\varphi_{-1}(\varphi_t(x_2))
\frac{u\circ\varphi_t(x_2)}{|u\circ\varphi_t(x_2)|}\right| &
=\Lambda .\end{split}\end{equation} Note that
$u\circ\varphi_t(x)=D\varphi_t(x)u(x)$ for all $t\in \mathbb{R}$
and $x\in \mathbb{T}^2$. Since $\| D\varphi_t(x)\|$ is bounded, it
suffices to prove the lemma for $t=k\in \mathbb{Z}$. If
$y_k=\varphi_k(x_1)$, $k=0,1,2,\ldots ,$ then
\begin{equation*}\begin{split} |u(y_k)|&=\left| D\varphi (y_{k-1})\frac{u(y_{k-1})}{{|u(y_{k-1})|}}\right| \cdot \left
|D\varphi (y_{k-2})\frac{u(y_{k-2})}{|u(y_{k-2})|}\right| \cdot \ldots\\ & \cdot \left| D\varphi
(y_1)\frac{u(y_1)}{|u(y_1)|}\right| \cdot \left|
D\varphi (y_0)\frac{u(y_0)}{|u(y_0)|}\right| \cdot |u(y_0)|
.\end{split}\end{equation*} Now \eqref{LambdaEst} implies \eqref{EstF} with $\epsilon <(\Lambda -\lambda )/2$. If
$y_k=\varphi_k(x_2)$, $k=0, -1,-2,\ldots $, then
\begin{equation*}\begin{split} |u(y_k)|&=|D\varphi_k(x_2)u(x_2)|\\ &=\left|
D\varphi_{-1}(y_{k+1})\frac{u(y_{k+1})}{|u(y_{k+1})|}\right| \cdot \left| D\varphi_{-1}(y_{k+2})
\frac{u(y_{k+2})}{|u(y_{k+2})|}\right|\cdot \ldots \\
&\cdot \left| D\varphi_{-1}(y_0)\frac{u(y_0)}{|u(y_0)|}\right|\cdot |u(y_0)|.\end{split}\end{equation*}
Now \eqref{LambdaEst} implies \eqref{EstB} with $\epsilon <(\lambda +\Lambda )/2$.
\end{proof}

\begin{proof}[Proof of Lemma~\ref{LemSL}] For $H$ defined in the Introduction, we have:
$$DH(t,\tau)=D\varphi_t(\psi _\tau x_0)\left[ u \vdots \frac{u^{\bot}}{|u|^2}\right]\circ \psi_\tau (x_0)$$
and $\det DH=1$. For $F$ and $\tilde{F}$, defined in \eqref{DefF} and \eqref{DefFTil}, we have:
$$\nabla F=\begin{bmatrix}\alpha F+e^{\alpha t}\gamma '(t)\beta (\tau )\\ e^{\alpha t}\gamma (t)\beta '(\tau
)\end{bmatrix};\quad \nabla \tilde{F}=\begin{bmatrix} \alpha \tilde{F}+e^{\alpha t}\gamma '' (t)\beta (\tau )\\ e^{\alpha
t}\gamma ' (t)\beta ' (\tau )\end{bmatrix}.$$
Choose $\gamma\in H_1([-N,N])$ and $\beta (\tau )=(s-|\tau |)\mathbb{I}_{[-s,s]}$, where $\mathbb{I}$ is the
characteristic function. If $s\to 0$ then $|\partial F/\partial t|^2/(2s)\to 0$ and $|\partial \tilde{F}/\partial
t|^2/(2s)\to 0$ in $L^2$. Also, $|\partial F/\partial \tau |^2/(2s)\to e^{2\lambda t}|\gamma (t)|^2\delta_0(\tau )$ and
$|\partial \tilde{F} /\partial \tau |^2/(2s)\to e^{2\lambda t}|\gamma '(t)|^2\delta_0 (\tau  )$ where $\delta_0(\cdot )$ is
the Dirac $\delta$-function. The measure of the support of $f$ tends to zero
as $s\to 0$, and $f/\| f\|_{H_1}\to 0$ weakly.
Since $L=-A+T$, where $T$ is a compact operator, there is a subsequence $s_j\to 0$
such that $\| T(f/\| f\|_{H_1})\| \to 0$.
Note that
$$[DH^{-1}(H(t,\tau ))]^\top=[D\varphi_t(\psi _\tau x_0)]^{-1\top}
\left[ \frac{u}{|u|^2}\vdots u^{\bot}\right]\circ \psi_\tau
(x_0).$$ Passing to the $(t,\tau )$ coordinates and letting $j\to \infty$ we have:
\begin{equation*}\begin{split} \| L+\alpha \|^2_{\bullet} 
&\lesssim \| Af-\alpha f\|^2_{H_1}/\| f\|^2_{H_1}+\| T(f/\|
f\|_{H_1})\|^2_{H_1}\\
&= \frac{(2s_j)^{-1}\int\limits_{\mathcal{S}}|DH^{-1\top}
\nabla \tilde{F}|^2d\tau dt}{(2s_j)^{-1}
\int\limits_{\mathcal{S}}|DH^{-1\top}\nabla
F|^2d\tau dt}+\| T(f/\| f\|_{H_1})\|^2_{H_1}\\
&\to \frac{\int\limits_{\mathbb{R}}|
[D\varphi_t(x_0)]^{-1\top}u^{\bot}(x_0)|^2e^{2\lambda t}|\gamma
'(t)|^2dt}{\int\limits_{\mathbb{R}}|[D\varphi_t(x_0)]^{-1\top}u^{\bot}(x_0)|^2e^{2\lambda t}(\gamma
(t)|^2dt}.\end{split}\end{equation*}
Since $[D\varphi_t(x_1)]^{-1\top}u^\bot (x_0)=u^\bot \circ \varphi_t(x_0)$,
this proves the lemma. To make $f$ mean-zero, define
another $\bar{f}$ in the same way around the same orbit and disjoint from $f$, varying its support we can obtain
$f-\bar{f}\in H^0_1$.
\end{proof}

\end{document}